\documentclass[12pt]{iopart}
\def\rref#1{(\ref{#1})}
\newcommand{\beq}{\begin{equation}}
\newcommand{\eeq}{\end{equation}}
\newcommand{\I}{{\hbox{{\rm I}\kern-.2em\hbox{\rm I}}}}
\newcommand{\bea}{\begin{eqnarray}}
\newcommand{\eea}{\end{eqnarray}}
\newcommand{\be}{\begin{equation}}
\newcommand{\ee}{\end{equation}}
\newcommand{\um}{{1\over 2}}
\newcommand{\ba}{\begin{array}}
\newcommand{\ea}{\end{array}}
\newcommand{\IR}{{\hbox{{\rm I}\kern-.2em\hbox{\rm R}}}}
\newcommand{\al}{\alpha}
\newcommand{\ga}{\gamma}
\newcommand{\sg}{\sigma}
\newcommand{\La}{\Lambda}
\newcommand{\n}{\eta}
\newcommand{\ep}{\epsilon}
\begin{document}

\title{Global constants in (2+1)--dimensional gravity}

\author{J E Nelson\footnote[1]{nelson@to.infn.it}}

\address{Dipartimento di Fisica Teorica, Universit\`a degli Studi di Torino and
Istituto Nazionale di Fisica Nucleare, Sezione di Torino, via Pietro Giuria 1,
10125 Torino, Italy}

\begin{abstract}  The extended conformal algebra $\hbox{so}(2,3)$ of global,
quantum, constants of motion in 2+1 dimensional gravity with topology $R \times
T^2$ and negative cosmological constant is reviewed. It is shown that the 10
global constants form a complete set by expressing them in terms of two
commuting spinors and the Dirac $\gamma$ matrices. The spinor components are
the globally constant holonomy parameters, and their respective spinor norms
are their quantum commutators.    
\end{abstract}



\maketitle

\section{Introduction}  Starting around 1989 there followed a period of intense
activity in the field of (2+1)--dimensional gravity, and a number of different
approaches were developed.  Some of these are the reduced phase space
quantization  with ADM variables \cite{vm1,hos,fuj}, quantization of the space
of classical solutions of the first-order Chern-Simons theory
\cite{Wit,Achu,car1}, and quantization of the holonomy algebra \cite{nr1}.  

In a pioneering article \cite{vm1} Vincent Moncrief reduced (\`a la ADM) the
Einstein equations for pure 2+1 gravity (with or without a cosmological
constant $\Lambda$, and for spacetimes with compact Cauchy surface), to a time 
dependent, finite dimensional, Hamiltonian system on the cotangent bundle of
the Teichm\"uller space of the Cauchy surface. At the same time Regge and I \cite{nr1} were studying the first order, holonomy formalism of
(2+1)--dimensional gravity, following Witten \cite{Wit} and Ach{\'u}carro  and
Townsend \cite{Achu}. In 1990 both Moncrief \cite{vm2} and Carlip \cite{car1}
started studying Cauchy surfaces diffeomorphic to the torus, since although it
is, in principle, possible to determine the evolution of the Teichm\"uller
parameters and their conjugate momenta (i.e., the reduced ADM variables), and
therefore solve Hamilton's equations implicitly, this procedure can be carried
out in practice only for Cauchy surfaces diffeomorphic to the torus, since the
spherical case is essentially trivial and the higher genus case particularly
unyielding. At about the same time Regge, Zertuche and I \cite{nrz} were
studying the case of non--zero cosmological constant, and Carlip had
determined \cite{car1} the relationship, for the case of the torus $T^2$,  and
with $\Lambda = 0$, between the reduced ADM variables and the holonomy
parameters, as a time--dependent canonical transformation. Meanwhile Moncrief
had found \cite{vm2} a set of six global constants of the motion, for the case
of zero cosmological constant, using a construction due to Martin \cite{mart}.
Linear combinations of these global constants satisfy the Lie algebra of the
Poincar\`e group $\hbox{ISO}(1,2)$. These constants were exactly the traces of
the  $\hbox{SO}(1,2)$ holonomies of the connections that Regge and I were
studying. 

A few years later Carlip and I \cite{cn1} had extended the analysis of the 
relationship (both classical and quantum) between ADM variables and holonomy
parameters, for the torus $T^2$, to $\Lambda \neq 0$, and it was natural to
apply this to Moncrief's global constants, since, although quantization in
terms of the ADM variables was studied in \cite{puz}, it involves non--trivial
operator ordering ambiguities and the square root of the Laplace--Beltrami
operator \cite{carord}.

Firstly, Moncrief and I extended the construction of global constants 
to non--zero (negative)
cosmological constant. This was straightforward \cite{mn}, and the six new
global constants were found to indeed reduce to those of \cite{vm2} in the
limit $\Lambda \to 0$, and to satisfy the Lie algebra of the anti--de Sitter
group $\hbox{SO}(2,2)$. Using the classical relationship found in \cite{cn1}
the constants are particularly simple in terms of the holonomy parameters, and
easily quantized \cite{mn}.

In a related classical construction \cite{haj} for zero cosmological constant
in the unreduced, ADM, Hamiltonian formalism it was shown that on inclusion of
the globally constant part of the Hamiltonian (the full ADM Hamiltonian is a
function of time times a global constant) three new global
constants are derived and the algebra extends to that of the conformal algebra 
$\hbox{so}(2,3)$, whose corresponding group is the conformal group of
3-dimensional Minkowski space. The same was true for $\Lambda \neq 0$ and three
new quantum constants were obtained, which, together with the Hamiltonian, form
a null $\hbox{SO}(2,2)$ vector \cite{mn}.

Here I show that there exactly no more global constants by expressing them in
terms of two two--component commuting constant spinors (whose components are
the globally constant holonomy parameters).

The plan of the paper is as follows. In section 2 the alternative classical 
and quantum descriptions, and the relationship between them, are discussed. In 
section 3 six global constants are constructed, and expressed either in
ADM variables, or the holonomy parameters. Quantization is straightforward in
terms of the holonomy parameters. In section 4 the extended algebra is
calculated. In section 5 it is shown that the set of constants is complete by
explicit use of a spinor representation, and in section 6 the role of the
modular group is briefly discussed. 

\section{Hamiltonian Dynamics}\label{subsec:prod} 
\subsection{Reduced ADM Dynamics}
It is known \cite{mn,cn1}
that  (2+1)-dimensional gravity, with topology $ R
\times T^2$, the York time coordinate condition, and with or without a 
cosmological constant $\Lambda$, has (at least) two equivalent descriptions. 
In ADM quantization 
the reduced Einstein action can be written \cite{vm1, hos}
\begin{equation}
I_1 = \int d\tau \left( p^\alpha {d{q_\alpha}\over {d\tau}} - H(q,p,\tau)
\right), ~~~\alpha = 1,2
\label{bb10}
\end{equation}
where the constant mean curvature $\tau$ labels the slices $T^2$,
and $q_1, q_2, p^1, p^2$ are the reduced ADM canonical coordinates and momenta.
 In equation \rref{bb10} the reduced, or effective, ADM  Hamiltonian $H$
is just the spatial volume
\be H = \int_{T^2} d^2x \sqrt{^{(2)}g} = {1 \over   \sqrt{\tau^2 - 4\Lambda}}
~\bar{H}, \qquad  \bar{H} = \sqrt{p_1^2 + e^{-2q^1}p_2^2}. \label{11} 
\ee 
and is not constant, since volume is not conserved, but is, instead, a function
of time times a global constant $\bar H$. From (\ref{bb10}) the basic Poisson
brackets are
\begin{equation}
\left\{ q_i, p^j \right\} = \delta _i{}^j
\label{bb14}
\end{equation}
As discussed in \cite{puz,carord} quantization is achieved by replacing equation 
\rref{bb14} with the commutators
\begin{equation}
\left[ {\hat q}_i, {\hat p}^j \right] = 
 i\hbar\delta_i{}^j ,
\label{da1}
\end{equation}
representing the momenta as derivatives,
\begin{equation}
p^j = {\hbar\over i}{\partial\ \over\partial q_j} ,
\label{da2}
\end{equation}
and imposing the Schr\"odinger equation
\begin{equation}
i\hbar{{\partial\psi(q,\tau)}\over{\partial \tau}} = \hat H\psi(q,\tau) ,
\label{da3}
\end{equation}
where the Hamiltonian $\hat H$ is obtained from (\ref{11}) by some
suitable operator ordering. The obvious choice is that of equation
(\ref{11}), for which the Hamiltonian becomes
\begin{equation}
\hat H = {\hbar\over\sqrt{\tau^2-4\Lambda}}\,\Delta_0^{1/2} ,
\label{da4}
\end{equation}
where $\Delta_0$ is the ordinary scalar Laplacian for the constant
negative curvature moduli space. Other orderings would correspond to operators 
$\Delta_n$, the weight $n$ Maass Laplacians \cite{maa}, which differ from $\Delta_0$
by terms of order $\hbar$. 

\subsection{The holonomy representation}
In the fully reduced holonomy representation \cite{nr1,nrz} the constraints 
are solved exactly, the Hamiltonian is zero, and (2+1)--dimensional gravity is
described, for the torus, by the real, global, time-independent 
traces $R_1^{\pm},R_2^{\pm}, R_{12}^{\pm}$ of $\hbox{SL}(2,{\bf R})$ holonomies
which satisfy the Poisson bracket algebra (cyclical in the three $\pm$ traces)
\be
\{R_1^{\pm},R_2^{\pm}\}=\mp{1\over 4\alpha}(R_{12}^{\pm}-
 R_1^{\pm}R_2^{\pm}). 
\label{pbr}
\ee
In equation \rref{pbr} the two $\pm$ copies refer to the decomposition of the
spinor group of $\hbox{SO}(2,2)$ (the anti-de Sitter group) as
$\hbox{SL}(2,{\bf R})\otimes \hbox{SL}(2,{\bf R})$ \cite{nr1}, and the
subscripts  1, 2  refer to two intersecting paths $\ga_1, \ga_2$ on $T^2$ with
intersection  number $+1$. The third holonomy,  $R^\pm_{12}$ corresponds to the
path $\ga_1\cdot\ga_2$, which has intersection  number $-1$ with  $\ga_1$ and
$+1$ with $\ga_2$, and the constant $\alpha$ is related to the cosmological
constant through $\La = -1/{\alpha}^2$. The holonomies of \rref{pbr} can
be parametrized as\footnote[7]{Direct quantization of the algebra \rref{pbr}
gives an algebra related to the Lie algebra of the quantum group
$\hbox{SU}(2)_q$ \cite{nrz,su}, where $q=\exp{(4i\theta)}, \tan\theta=
- \hbar/{8\alpha}$. A (scaled) representation of the operators \rref{rpar} leads to
the commutators $[\hat r_1^{\pm}, \hat r_2^{\pm}] = \pm 8i\theta$, which differ
from \rref{dc1} by terms of order $\hbar^3$.}
\be
\eqalign{R_1^{\pm} = \cosh{r_1^{\pm} \over 2}\cr
R_2^{\pm} = \cosh{r_2^{\pm} \over 2}\cr
R_{12}^{\pm} = \cosh{(r_1^{\pm} + r_2^{\pm})/2}}
\label{rpar}
\ee 
where
$r_{1,2}^{\pm}$ are also real, global, time-independent (but undetermined) 
parameters which, from equation \rref{pbr} satisfy
\be\{r_1^\pm,r_2^\pm\}=\mp {1\over\al}, \qquad \{r_{1,2}^+,r_{1,2}^-\}= 0.
\label{pb}
\ee
With this parametrisation, the Chern-Simons action \cite{Wit, Achu} is \cite{cn1}
\begin{equation}
I_2 = \int \alpha(r_1^-dr_2^- - r_1^+ dr_2^+)
\label{i2}
\end{equation}
and quantization is achieved by replacing equations \rref{pb} with the commutators
\begin{equation}
[\hat r_1^\pm, \hat r_2^\pm] = \mp {i\hbar/\alpha}, \quad [\hat r_{1,2}^+, 
\hat r_{1,2}^-] = 0 .
\label{dc1}
\end{equation}
 
\subsection{Relationship between the ADM and holonomy descriptions}
The above two descriptions appear quite different in concept and in structure,
but classically can be related because of the availability of explicit 
classical solutions for the torus case. The conformal spatial metric 
$g^{-1/2} g_{ij}$ is in fact parametrized by the parameters $q^1, q^2$ 
\be 
g^{-1/2} g_{ij}= \left(\matrix{e^{-q^1}+(q^2)^2e^{q^1}&q^2e^{q^1}\cr
q^2e^{q^1}&e^{q^1}\cr }\right)
\label{sol}
\ee
Now, for a suitable triad $e^a=e^a{}_{\mu}dx^{\mu}$ whose spatial components $e^a_{i}$ 
are related to the two metric through
\be
g_{ij} = e^a_{i} e^b_{j}\n_{ab}, \quad \n_{ab} = diag (-1,1,1), 
\quad a,b,c = 0,1,2,  
\ee
and its corresponding spin connection 
\be 
\omega^{a} = {1 \over 2} ~\epsilon^{abc} \omega_{bc},
\quad \ep^{012}= - \ep_{012}=1 
\ee
the traces of holonomies $R_i^{\pm}$ (equations \rref{rpar}) are just the 
Wilson loops 
\be R_i^{\pm}= Tr \left( \exp \Delta_i^{\pm}{}^a \right),~~i=1,2,12
\label{tr}
\ee
where 
\be
\Delta_i^{\pm}{}^a = \int_{\ga_i} \lambda^{\pm}{}^{(a)}
\label{del}
\ee
and the  $\lambda^{\pm}{}^{a}$ are "shifted connections" 
defined by 
\be
\lambda^{\pm}{}^{a} = \omega^{a} \pm \sqrt{- \Lambda} ~e^{a} .   
\label{lam}\ee 
These integrated shifted connections \rref{del} and therefore the 
traces \rref{tr} can be calculated directly from the classical solutions 
\rref{sol}, and their relation to the parametrisation \rref{rpar} is 
\be
(r_{1,2}^{\pm})^2= \Delta_{1,2}^{\pm}{}^a~\Delta_{1,2}^{\pm}{}^b~\eta_{ab},
\ee
From equations \rref{sol}--\rref{lam} we have the following \cite{cn1}
\bea
m & = &\left(r_1^-e^{it/\al} + r_1^+e^{-{it/\al}}\right) 
 \left(r_2^-e^{it/\al} + r_2^+e^{-{it/\al}}\right)^{\lower2pt%
 \hbox{$\scriptstyle -1$}}, \nonumber \\
\pi & = & -{i\al\over 2\sin{2t\over \al}}\left(r_2^+e^{it/\al} 
 + r_2^-e^{-{it/\al}}\right)^{\lower2pt%
 \hbox{$\scriptstyle 2$}}, 
\label{13}
\eea
where, in equation \rref{13} $m = m_1 +im_2$ are the complex moduli, 
and $\pi=\pi^1 +i \pi^2$ their complex momenta, related to the ADM coordinates
and momenta through
\be
m_1 = q^2, m_2 = e ^{-q^1}, \pi^1 = p_2, \pi^2 = -p_1 
e^{q^1}  
\label{14}
\ee
and the mean curvature $\tau$ is related to $t$ in equations \rref{13}--
\rref{14} by $ \tau  = -{2\over\alpha}\cot{2t\over\alpha} $, 
and is monotonic in the range $t~ \epsilon~ (0, {\pi \al / 2})$. Note that 
since the $r_1^{\pm}, r_2^{\pm}$ are arbitrary the moduli and momenta 
of equation \rref{13} (and therefore, from equation \rref{14}, also the ADM
variables) can have arbitrary initial data at some initial time $t_0$.

That the two descriptions are related classically through a time-dependent 
canonical transformation can be seen, from equations \rref{13}--\rref{14}, 
from the comparison of the classical actions \rref{bb10} and \rref{i2} 
\begin{equation}
I_1 = I_2 + \int d(\pi^1 m_1 +\pi^2 m_2) = I_2 + \int d(p_2 q^2 - p_1)
\end{equation}
Using \rref{13} and \rref{14} the ADM Hamiltonian \rref{11} becomes 
\be
H = {\al \over {2\sqrt{\tau^2 - 4\Lambda}}}(r_1^-r_2^+ 
- r_1^+r_2^-),
\quad \bar H = {\al \over 2} ( r_1^- r_2^+ 
-  r_1^+ r_2^-).
\label{15}
\ee
and quantum mechanically, there would be no ordering ambiguity, as can be seen
from equation \rref{dc1}. It should be noted, however, that no representation
is known for the fundamental quantized holonomy parameters $r_1^{\pm},
r_2^{\pm}$ which guarantees positivity of the reduced Hamiltonian \rref{15},
though some progress has been made using the representation of \cite{cn2}. 

Comparing the two quantum theories is much more subtle, and I will say very
little. The ADM representation looks like a standard Schr{\"o}dinger picture
quantum theory, with time-dependent states $\psi(q,\tau)$ whose evolution is
determined by a Hamiltonian operator, equation \rref{da4}.  The holonomy
representation resembles a Heisenberg picture quantum theory, characterized by
time-independent states $\psi(r_1^{\pm}, r_2^{\pm})$ (though clearly some
choice of polarization would be necessary) and time-dependent operators
\rref{13}--\rref{14}. A polarization and unitary transformation between the 
two representations was constructed in \cite{cn2}.

\section{Constants of the motion}  It is known \cite{mart} that the traces of
$\hbox{SO}(1,2)$ holonomies, for $\Lambda = 0$, of connections integrated along
arbitrary closed loops, are absolutely conserved quantities
(i.e., they are gauge invariant and invariant under non-singular deformations
of the loops within the vacuum spacetimes).

Here for $\Lambda < 0$ it is instead appropriate to use the integrated shifted
connections \rref{del} for  $\hbox{SL}(2,{\bf R}) \otimes \hbox{SL}(2,{\bf
R})$, the spinor decomposition of  $\hbox{SO}(2, 2)$, the anti--de Sitter
group. As in \cite{vm2} these traces were computed for 3-different classes of
loops corresponding to the paths $\ga_1, \ga_2$ and "twisting loops"
$\ga_1\cdot\ga_2$. In spatial coordinates $x^1, x^2$ these loops would
correspond to having $x^2 = constant$, $x^1 = constant$, whereas the "twisting
loops" have $x^1 = {m \over n} ~x^2$, for integer $m,n$, respectively. The
twisting loops do not give new independent conserved quantities but, instead,
are functions of those coming from the loops $\ga_1$ and $\ga_2$.

\subsection{ADM Variables}

The following absolutely conserved quantities were obtained \cite{mn}, as in 
\cite{vm2,mart}, from the traces of the  $\hbox{SL(2,\bf R)}$ holonomies 
\bea
&C_1^{\pm}& =C_1 \pm 2\sqrt{-\Lambda} C_4, \nonumber \\
&C_2^{\pm}& =C_2 \mp 2\sqrt{-\Lambda} C_5, \nonumber \\
&C_3^{\pm}& =C_3 \pm \sqrt{-\Lambda} C_6,
\label{21}
\eea
where
\begin{eqnarray}
&C_1 &= {1 \over 2} ~e^{-q^1} \tau \left\{ (\sqrt{1 - {4\Lambda \over  
\tau^2}} \bar{H} - p_1)(1 + (q^2)^2 e^{2q^1}) - 2(q^2 p_2 - p_1)  
\right\} , \nonumber \\
&C_2 &= {1 \over 2} ~e^{q^1} \tau\left\{\sqrt{1 - {4\Lambda \over  
\tau^2}} \bar{H} - p_1 \right\} , \nonumber \\
&C_3 &= {1 \over 2} ~e^{q^1} \tau\left\{ q^2 (\sqrt{1 - {4\Lambda  
\over \tau^2}} \bar{H} - p_1) - p_2  ~e^{-2q^1} \right\} , \nonumber \\
&C_4 &= {1 \over 2} \left\{ p_2 ~e^{- 2q^1} + 2 q^2 p_1 - p_2 (q^2)^2  
\right\} , \nonumber \\
&C_5 &= {1 \over 2} ~p_2, \nonumber \\
&C_6 &= p_1 - q^2 p_2.
\label{22}
\end{eqnarray}
The above quantities $C_1 - C_6$ are {\it quasi--linear} in the momenta 
$p_1, p_2$, as is the
ADM Hamiltonian itself (see equation \rref{11}), and in the limit 
$\Lambda \rightarrow 0$ reduce to those defined in \cite{vm2}.
That they are conserved quantities can be verified directly through  
Hamilton's equations (where $H$ is given by equation \rref{11})
\be
{dq^i \over {d \tau}} = {{\partial H}  \over {\partial p_i}}  
~,~~ {dp_i \over {d \tau}} = - {{\partial H} \over {\partial  
q^i}}, \quad i = 1,2.  
\ee
The $C_1 - C_6$ are related as follows to the perhaps more 
familiar quantities $J_{ab}, P_c$ through
\be
\eqalign{P_0&= - \um (C_1 + C_2),\\ P_1&=\um (C_1 - C_2),\\
  P_2&= C_3, \\
J_{12}&= C_5 - C_4,\\
 J_{02}&= C_4 + C_5,\\ J_{01}&= - C_6 } 
\label{pjs}\ee 
which, from equation \rref{bb14}, satisfy the Lie algebra $\hbox{so}(2,2)$ 
of the anti--de Sitter group
\be
\eqalign{\{J_{ab}, J_{cd}\}&= \n_{ac} J_{bd} - \n_{bc} J_{ad} -\n_{ad} 
J_{bc}+ \n_{bd} J_{ac}\\
 \{P_a , P_b\}&= \La J_{ab},\\
\{J_{ab}, P_c\}&= \n_{ac}P_b - \n_{bc}P_a}
\label{pjalg}\ee
with the identity
\be P_a J_{bc} \ep^{abc} = C_4 C_2 - C_5 C_1 - C_3 C_6 = 0,   
\label{pj}
\ee
The constant part $\bar H$ of the ADM Hamiltonian \rref{11}
is related to these quantities through
\be \eqalign{\Lambda {\bar H}^2 &= (C_3)^2 - C_1 C_2 = P_a P^a\\
{\bar H}^2 &= (C_6)^2 + 4 C_4 C_5 = - J_{ab} J^{ab}}
\label{h2}
\ee
Quantization of these constants and the Hamiltonian (equation \rref{11}) 
in terms of these ADM variables has been discussed in \cite{puz}.
\subsection{Holonomy parameters}
In terms of the time independent global parameters $r_{1,2}^{\pm}$ 
the $C_1^{\pm}  - C_3^{\pm}$, equation \rref{21} are, from equations 
\rref{13}--\rref{14}, quite simple, and evidently globally conserved
\be
C_1^{\pm} = (r_1^{\mp})^2, \quad C_2^{\pm} = (r_2^{\mp})^2, \quad
C_3^{\pm} = r_1^{\mp}r_2^{\mp},
\label{23}
\ee 
and quantization is indeed straightforward 
in terms of these parameters. From equations \rref{21}, \rref{22} and \rref{pjs} 
the combinations
\be
{\hat j}_a^{\pm}=  {1\over 2} \epsilon_{abc}J^{bc} \pm \alpha P_a  
\label{j}
\ee
are just
\be
{\hat j}_0^{\pm} = \mp {\al \over 2} ({({\hat r}_1^{\pm})}^2 
+ {({\hat r}_2^{\pm})}^2),
\label{j0}
\ee
\be
{\hat j}_1^{\pm} = \pm {\al \over 2} ({({\hat r}_1^{\pm})}^2 
- {({\hat r}_2^{\pm})}^2), 
\label{j1}
\ee
\be
{\hat j}_2^{\pm} = \pm {\al\over 2}
({\hat r}_1^{\pm}{\hat r}_2^{\pm} + {\hat r}_2^{\pm}{\hat r}_1^{\pm}).   
\label{j2}
\ee
Note that the $j_a^+$ depend only on the $r^+$'s and the $j_a^-$ 
only on the $r^-$'s, and that the only ordering ambiguity is in 
${{\hat j}_2}{}^{\pm} $, equation \rref{j2}. With this symmetric ordering 
the combinations \rref{j} satisfy, using equations \rref{dc1}, the two 
$(\pm)$ Lie algebras of $\hbox{so}(1,2) \approx \hbox{sl(2, \bf R)}$
\be
[{\hat j}_a^{\pm},{\hat j}_b^{\pm}] = 2i {\hbar} \epsilon_{abc}
{{\hat j}^c}{}^{\pm},  \quad [{{\hat j}_a}^+,{{\hat j}_b}^-] = 0,  
\label{26}
\ee
Now the generators ${\hat j}_a^{\pm}$ and $\hat {\bar H}$ are not all 
independent. There are 3 Casimirs corresponding to the classical 
identities \rref{pj} and \rref{h2}
\be
{\hat j} = {{\hat j}_a}^{\pm} {{\hat j}^a}{}^{\pm}= {3{\hbar}^2 \over  
4},  \quad {\hat {\bar H}}^2 - \um{{\hat j}_a}^+ {{\hat j}^a}{}^- 
= {{\hbar}^2 \over 2}.   
\label{27}
\ee
Note that for the only ordering ambiguity is in ${{\hat j}_2}{}^{\pm} $ 
(equation \rref{j2}) any other ordering would only produce terms of 
$O(\hbar^2)$ on the right hand side of equation \rref{26} and equation \rref{27}.
\section{The Extended Algebra}
In a related classical construction \cite{haj} for zero cosmological constant
the constants $C_1 - C_6$ \rref{22} were used as generators 
of isometries in the unreduced, ADM, Hamiltonian formalism. Consider the
constant part $\hat {\bar H}$ (equation \rref{15}) of the ADM Hamiltonian.
It can be checked, using  equation \rref{dc1}, that
$\hat {\bar H}$ does not commute with all the 
$\hbox{sl(2, \bf R)}$ generators, equations \rref{j0}--\rref{j2} (though it does
commute with ${\hat j}_a^+ + {\hat j}_a^- = \epsilon_{abc}J^{bc} $). Since
\be
[\hat {\bar H},{\hat r}_i^{\pm}] = i {\hbar\over 2} {\hat r}_i^{\mp}, 
\ee
a new globally constant three-vector ${\hat v}_a$ is defined through
\be
[\hat {\bar H},{\hat j}_a^{\pm}] = \pm i {\hbar} {\hat v}_a, 
\label{31}
\ee
where
\bea
{\hat v}_0&=&-{\al \over 2}({\hat r}_1^+ {\hat r}_1^- + {\hat r}_2^+ 
{\hat r}_2^-),\cr 
{\hat v}_1&=&~~{\al \over 2}({\hat r}_1^+ {\hat r}_1^- - {\hat r}_2^+ 
{\hat r}_2^-), \cr
{\hat v}_2&=&-{\al \over 2}({\hat r}_1^+ {\hat r}_2^- + 
{\hat r}_2^+ {\hat r}_1^-).
\label{v}
\eea
Note that from \rref{dc1} there are also no ordering ambiguities in the 
${\hat v}_a$.

The extended algebra of the {\it ten} $\hat {\bar H}, {\hat j}_a^{\pm},  {\hat
v}_a, a=0,1,2$ (equations \rref{15}, \rref{j0}--\rref{j2}, \rref{v}),  then
closes as follows  
\bea 
[\hat {\bar H},{\hat j}_a^{\pm}] &=& \pm i {\hbar} {\hat
v}_a \cr
[{\hat j}_a^{\pm},{\hat j}_b^{\pm}] &=&~~ 2i {\hbar} \epsilon_{abc}
{{\hat j}^c}{}^{\pm} \cr
[{{\hat j}_a}^+,{{\hat j}_b}^-] &=& ~~0 \cr
[\hat{\bar H},{\hat v}_a] &=&~ {i\hbar \over 2}({\hat j}_a^+ - {\hat j}_a^-) \cr 
[{\hat v}_a,{\hat v}_b] &=& -{i \hbar \over 2} \epsilon_{abc} ({{\hat j}^c}{}^+
+ {{\hat j}^c}{}^-) \cr 
[{\hat j}_a^{\pm},{\hat v}_b] &=& ~i \hbar (\mp 
\eta_{ab} \hat {\bar H}  + \epsilon_{abc} {\hat v}^c) \label{alg}
\eea 
with the
3 additional identities (making, with equation \rref{27}, a total of 6
identities) 
\be {\hat v}^a{\hat j}_a^{\pm} = {\hat j}_a^{\mp}{\hat v}^a =  \pm
{3i\hbar \over 2}\hat {\bar H}, and \label{34a} 
\ee 
\be 
{\hat v}_a{\hat v}^a -
{\hat {\bar H}}^2 = - {\hbar^2 \over 2}. \label{34} 
\ee  
The above
10-dimensional algebra \rref{alg} is isomorphic to the Lie algebra  of
$\hbox{so}(2,3)$, whose corresponding group is the conformal group of 
3-dimensional Minkowski space \cite{haj}. The dilatation $D$ is to be 
identified with - $ \hat {\bar H}$, the translations with ${\hat P}_a^-$, and
the conformal accelerations with ${\hat P}_a^+$, where ${\hat P}_a^{\pm} =
\alpha {\hat P}_a \pm {\hat v}_a$. Note that, in contrast  to the generators
${\hat j}_a^+$ and ${\hat j}_a^-$ (equations  \rref{j0}--\rref{j2}) of the two
commuting $\hbox{sl(2, \bf R)}$ subalgebras  (equation \rref{26}), the group
extension, that is, the Hamiltonian  $\hat {\bar H}$ (equation \rref{15} and
the vectors ${\hat v}_a$ (equation \rref{v}) require {\it both} the ${\pm}$
global  parameters, and therefore {\it both} the two $\hbox{sl}(2, \bf R)$
algebras.  
\section{Spinor representation}\label{spinor} The fact that the
Hamiltonian $\hat {\bar H}$ and the three new constants  ${\hat v}_a$ define a
null  $\hbox{SO}(2,2)$ vector (see equation \rref{34}) suggests the use of the 
{\it commuting} two--component constant spinors ${\hat r}^+,{\hat r}^-$ 
defined by 
\beq {\hat r}^{\pm} = {{\hat r}_1^{\pm}\choose {\hat r}_2^{\pm}}
\label{spin}
\eeq
which satisfy, from \rref{dc1}  $[{\hat r}^+, {\hat r}^-] = 0 $, and whose 
respective norms are just the commutators
\be
\ep^{AB}{\hat r}_B^{\pm}{\hat r}_A^{\pm}={{\hat r}^A}{}^{\pm}
{\hat r}_A^{\pm}= [{\hat r}_2^{\pm}, {\hat r}_1^{\pm}] = 
\mp {i {\hbar} \over \al}, \quad A,B = 1,2
\ee
with $\ep ^{12} = - \ep ^{21} = 1$ and 
${\hat r}^A{}^{\pm}=\ep ^{AB}{\hat r}_B^{\pm}$. 

In terms of these {\it commuting} two--component constant spinors 
${\hat r}^+,{\hat r}^-$
the ten constants $\hat {\bar H}, {\hat j}_a^{\pm}, {\hat v}_a$ are
\beq {\hat v}_0 = - {\al\over 2} {{\hat r}^+}{}^T \I {\hat r}^- \eeq
\beq {\hat v}_1 = ~~ {\al\over 2} {{\hat r}^+}{}^T \sg_3 {\hat r}^-\eeq
\beq {\hat v}_2 = ~~ {\al\over 2} {{\hat r}^+}{}^T \sg_1 {\hat r}^- \eeq
\beq \hat {\bar H} = - i {\al\over 2} {{\hat r}^+}{}^T \sg_2 {\hat r}^- \eeq

\beq \hat {j_0}^{\pm} = \mp {\al\over 2} {{\hat r}^{\pm}}{}^T \I {\hat
r}^{\pm} \eeq

\beq \hat {j_1}^{\pm} = \pm {\al\over 2} {{\hat r}^{\pm}}{}^T \sg_3 {\hat
r}^{\pm} \eeq

\beq \hat {j_2}^{\pm} = \pm {\al\over 2} {{\hat r}^{\pm}}{}^T \sg_1 {\hat
r}^{\pm} 
\eeq
where the $\sg_1, \sg_2, \sg_3$ are the usual Pauli matrices.

That there are precisely no more than these ten constants can be seen by 
writing these constants in terms of the Dirac $\ga$ matrices satisfying 
\beq \{\ga_a,\ga_b\}=2 \eta_{ab} 
\eeq
with $\eta_{ab} = diag(-1,1,1-1)$ the $\hbox{SO}(2,2)$ (anti-de Sitter) 
metric, and the {\it four}--component constant spinors
\beq {\hat r} = {{\hat r}^+\choose {\hat r}^-}
%= {{\hat r}_1^+\choose {\hat r}_2^+\choose {\hat r}_1^-\choose {\hat r}_2^-}
\eeq
with ${\hat r}^+,{\hat r}^-$ given in equation \rref{spin}. 

Consider the sixteen linearly independent $4 \times 4$ matrices $\Gamma_i,
i=1....16$ 
\be
\Gamma_i = \I, \ga_a, \ga_5 = i \ga_0\ga_1\ga_2\ga_3, \ga_a\ga_5, \ga_a\ga_b, 
a\neq b. 
\ee
There should therefore be at most sixteen global constants of 
the form 
\be{{\hat r}}{}^T \Gamma_i {\hat r}\label{gam}.\ee
Consider the representation \cite{nrz}
\beq \gamma_0 = \left(\matrix { &1& & \cr-1& & & \cr & & &-1 \cr & &1& \cr}\right)
= i \sigma_2\otimes\sigma_3=i \left(\ba{clcr}\sg_2& \\ &-\sg_2\ea\right)\eeq
\beq \gamma_1 = \left(\matrix {1& & & \cr &-1& & \cr & &-1 & \cr & & &1\cr}\right)
= \sigma_3\otimes\sigma_3= \left(\ba{clcr}\sg_3& \\ &-\sg_3\ea\right)\eeq
\beq \gamma_2 = \left(\matrix { & &-1& \cr & & &-1 \cr-1& & & \cr &-1& & \cr}\right)
= -\I\otimes\sigma_1=\left(\ba{clcr} &-\I \\ -\I& \ea\right)\eeq
\beq \gamma_3 = \left(\matrix { &-i& & \cr -i& & & \cr & & &i \cr & &i& \cr}\right)
= -i \sigma_1\otimes\sigma_3=i \left(\ba{clcr}-\sg_1& \\ &\sg_1\ea\right)\eeq
with
\beq \gamma_5 = i \ga_0\ga_1\ga_2\ga_3=\I\otimes\sg_2=i\left(\ba{clcr} &-\I \\
\I& \ea\right).\eeq

With this representation then of the sixteen possible global constants, 
equation \rref{gam}, five are identically zero, and one a numerical constant,
due to the commutators \rref{dc1}. The remaining ten are linear
combinations of the ten global constants $\hat {\bar H}, {\hat v}_i,{\hat
j}_i{}^{\pm}, i=0,1,2 $, as I shall show.

There are eight block diagonal combinations, namely 
$\ga_0,\ga_1,\ga_3,\I$ and

\beq \ga_0\ga_1 = -\sg_1\otimes\I=-\left(\ba{clcr} \sg_1& \\ &\sg_1\ea\right)\eeq 
\beq \ga_0\ga_3= -i \sg_3\otimes\I=-i\left(\ba{clcr} \sg_3& \\ &\sg_3\ea\right)\eeq
\beq \ga_1\ga_3=\sg_2\otimes\I=i\left(\ba{clcr} \sg_2& \\ &\sg_2 \ea\right)\eeq
\beq \ga_2\ga_5 = -i \I\otimes\sg_3=i\left(\ba{clcr} -\I& \\ &\I \ea\right)\eeq
and six of them combine to give the six constants $\hat {j}_a^{\pm}$ as 
\beq \hat {j_0}^{\pm} = \pm {\al\over 4} {{\hat r}}{}^T (\I\pm i \ga_2\ga_5)
{\hat r} 
\eeq
\beq \hat {j_1}^{\pm} = {\al\over 4} {{\hat r}}{}^T (\ga_1 \pm i \ga_0\ga_3)
{\hat r} 
\eeq
\beq \hat {j_2}^{\pm} = \mp {\al\over 4} {{\hat r}}{}^T (\ga_0\ga_1 \mp i \ga_3)
{\hat r}
\eeq
The two unused block--diagonal elements, namely $\ga_0$ and $\ga_1\ga_3$ do not
give other global constants, since 
\beq {{\hat r}}{}^T \ga_0 {\hat r}=i({{\hat r}^+}{}^T \sg_2 {\hat
r}^+ - {{\hat r}^-}{}^T \sg_2 {\hat
r}^-) = [{\hat r}_1^+,{\hat r}_2^+]-[{\hat r}_1^-,{\hat r}_2^-]= 
- {2i \hbar\over\al} \eeq 
and
\beq {{\hat r}}{}^T \ga_1\ga_3 {\hat r}={{\hat r}^+}{}^T \sg_2 {\hat
r}^+ + {{\hat r}^-}{}^T \sg_2 {\hat r}^- = -i([{\hat r}_1^+,{\hat r}_2^+]+
[{\hat r}_1^-,{\hat r}_2^-])= 0 \eeq 

The remaining eight possible combinations are all block off--diagonal, with
four symmetric, and four anti--symmetric, so the available constants are all 
of the form
\beq \left({\hat r}^+ ~{\hat r}^-\right)\left(\ba{clcr} & c\\d& \ea\right){{\hat r}^+
\choose {\hat r}^-}={\hat r}^- d {\hat r}^+ + {\hat r}^+ c {\hat r}^- \label{od}
\eeq
for some $2 \times 2$ matrices $c$ and $d$. For the four anti--symmetric
matrices  $\ga_1\ga_2,\ga_2\ga_3,\ga_0\ga_5$  and $\ga_5$ (i.e. with $c{}^T = -
d$) the two terms in \rref{od} cancel since $[{\hat r}^+, {\hat r}^-]=0$. The  
four symmetric matrices  $\ga_0\ga_2,\ga_2,\ga_1\ga_5,\ga_3\ga_5$ (i.e. with
$c{}^T = d$) instead combine to give the four constants $\hat {\bar H}$ and
${\hat v}_a$ respectively, as follows.
\beq \hat {\bar H} =  {\al\over 4} {{\hat r}}{}^T \ga_0\ga_2 {\hat r} \eeq
\beq {\hat v}_0 =  {\al\over 4} {{\hat r}}{}^T \ga_2 {\hat r} \eeq
\beq {\hat v}_1 = i {\al\over 4} {{\hat r}}{}^T \ga_1\ga_5 {\hat r}\eeq
\beq {\hat v}_2 = - {\al\over 4} {{\hat r}}{}^T \ga_3\ga_5 {\hat r} \eeq

\section{The Quantum Modular Group} So far I have not mentioned the role of the
"large diffeomorphisms" - diffeomorphisms that cannot be continuously deformed
to the identity. For the torus $T^2$ this group---also known as the modular
group---acts on the torus modulus and momentum and holonomy parameters as
\begin{eqnarray}
&S&: m\rightarrow -m^{-1} ,\quad
     \pi\rightarrow \bar m^2 \pi,
\quad r_1^{\pm}\rightarrow r_2^{\pm},\quad 
    r_2^{\pm}\rightarrow - r_1^{\pm},\nonumber\\
&T&: m\rightarrow m+1 ,\quad \pi\rightarrow \pi ,
\quad r_1^{\pm}\rightarrow r_1^{\pm} + r_2^{\pm},\quad  
    r_2^{\pm}\rightarrow r_2^{\pm},
\label{28}
\end{eqnarray}
and generates the entire group of large diffeomorphisms of 
$R \times T^2$. It can be seen from equation \rref{28} that this group  
acts properly discontinuously on the ADM variables, (the configuration
or Teichm{\"u}ller space variables, see equation \rref{14}) and in fact splits this space into
fundamental regions that are interchanged by the action of the group.

But on the holonomy parameters the group is {\it not} so well behaved. Even so,
with the ordering of equation \rref{13} (the only ambiguity), the quantum 
action of the modular group is the same as the classical one, with no
$O(\hbar)$ corrections, and is generated by the $\hbox{SO}(2,2)$ discrete
anti--de Sitter subgroup of the conformal group $\hbox{SO}(2,3)$ by 
conjugation with the operators $U_T = U_T^+ U_T^-,$  and $U_S = U_S^+ U_S^-$  
\cite{cn2,nr3} where 
\bea
U_T^{\pm} &=& \exp {{i \over 2\hbar}(j_0^{\pm} +j_1^{\pm})}= 
\exp {\mp{i\al \over 2\hbar}C_2^{\mp}} 
= \exp {\mp {i\al \over 2\hbar}(r_2^{\pm})^2},\nonumber \\
U_S^{\pm} &=& \exp {{i\pi \over 2\hbar}j_0^{\pm}}= 
\exp {\mp{i\pi\al \over 4\hbar}(C_1^{\mp} + C_2^{\mp})} 
= \exp {\mp {i\pi\al \over 4\hbar}((r_1^{\pm})^2) 
+(r_2^{\pm})^2)}.
\label{29}
\eea

Moreover, it is easy to check that the commutators \rref{dc1} and the ADM
Hamiltonian \rref{15} are invariant under the transformations \rref{28}. The
initial constants ${\hat j}_a^{\pm}$ transform into linear combinations of
themselves, as do the group extensions ${\hat v}_a$, in such a way that the
quantum algebra (equation \rref{26}) and the identities (equation \rref{27})
are invariant under the transformations \rref{28}.

\section{Conclusion} It is shown that there are exactly ten, and no more,
global constants satisfying the Lie algebra of $\hbox{so}(2,3)$, whose
corresponding group is the conformal group of 3-dimensional Minkowski space.
This is achieved by using two commuting two--component constant spinors whose 
norms are their respective commutators. With the six central elements
(equations \rref{27}, \rref{34a} and \rref{34}) this leaves precisely four
arbitrary global variables corresponding to the required two degrees of
freedom. These ten constants are easily expressed and quantized in terms of the
holonomy parameters.

It would be desirable to try to implement the operator analogues of 
$C_1 - C_6$, or alternatively ${\hat j}_a^{\pm}$ on a partially reduced quantization,
working on Teichm\"uller space for the torus (i.e. the 2-dimensional hyperbolic
space with global coordinates $q^1,  q^2 $ and Riemannian metric $(dq^1)^2 +
e^{2q^1}(dq^2)^2 $) instead of moduli space, but unfortunately there is no
known ordering of the operator analogues of expressions \rref{22} which
reproduces the $\hbox{so}(2,2)$ algebra as would be expected from classical
considerations.

It would also be desirable to be able to express the three-vector ${\hat v}_a$
(equation \rref{v}) in terms of the ADM variables. This is possible in
principle, from its definition \rref{31} but unfortunately the ${\hat v}_a$
and $\hat {\bar H}$ (equation \rref{15}) require {\it both } the ${\pm}$
global parameters. But, as can be seen from equation \rref{23} that would
imply that the ${\hat v}_a$ should be expressed as a square roots of quadratic
combinations of the $C_1 - C_6$. This is no surprise, since $\hat {\bar H}$ is,
from \rref{11}, or from \rref{h2} itself a square root of ADM variables, and is
related to the ${\hat v}_a$ through equation \rref{34}.

For completeness it should be noted that although only the case of negative
cosmological constant was discussed there would seem to be no obstruction  to
the discussion for $\Lambda$ positive or zero (see the discussion in
\cite{cn1}). For example, for $\Lambda > 0$, the parameters $r_1^{\pm},
r_2^{\pm}$  would be unchanged but the "shifted connections"  \rref{lam}, and
in consequence, the holonomies \rref{rpar} and the corresponding de Sitter or
$\hbox{sl(2,\bf C)}$  generators ${j_a}^{\pm}, a=0,1,2$, equation \rref{j}
would be complex conjugates of each other rather than real and independent as
here, for $\Lambda < 0$.

\vskip 0.7truecm 
\section*{Acknowledgments} This work was supported in part  
by INFN Iniziativa Specifica FI41, and by the MIUR under contract PRIN
2001-025492.

\end{document}